\documentclass[english,aps,11pt]{revtex4}
\usepackage{float}
\usepackage{graphicx}
\usepackage{algorithm}
\usepackage{algorithmic}

\makeatletter

\providecommand{\LyX}{L\kern-.1667em\lower.25em\hbox{Y}\kern-.125emX\@}
\floatstyle{ruled}

\@ifundefined{textcolor}{}
{%
 \definecolor{BLACK}{gray}{0}
 \definecolor{WHITE}{gray}{1}
 \definecolor{RED}{rgb}{1,0,0}
 \definecolor{GREEN}{rgb}{0,1,0}
 \definecolor{BLUE}{rgb}{0,0,1}
 \definecolor{CYAN}{cmyk}{1,0,0,0}
 \definecolor{MAGENTA}{cmyk}{0,1,0,0}
 \definecolor{YELLOW}{cmyk}{0,0,1,0}
 }

\renewcommand\[{\begin{equation}}
\renewcommand\]{\end{equation}}

\makeatother

\usepackage{babel}

\begin{document}

\preprint{This line only printed with preprint option}

\title{Multi-Determinant Wave-functions in Quantum Monte Carlo}

\author{Miguel A. Morales}
\email{moralessilva2@llnl.gov}
\affiliation{Lawrence Livermore National Laboratory, 7000 East Avenue, Livermore, California 94550, U.S.A.}

\author{Jeremy McMinis}

\affiliation{Department of Physics, University of Illinois at Urbana Champaign, Urbana, IL 61801}

\author{Bryan K. Clark}


\affiliation{Princeton Center For Theoretical Science, Princeton University, Princeton,
NJ 08544\\
Department of Physics, Joseph Henry Laboratories, Princeton University,
Princeton, NJ 08544}

\author{Jeongnim Kim}
\affiliation{National Center for Supercomputing Applications, University of Illinois at Urbana Champaign, Urbana, IL 61801\\
Materials Science and Technology Division and Computational Chemistry and Materials Division,
Oak Ridge National Laboratory, Oak Ridge, TN 37830, USA}

\author{Gustavo E. Scuseria}
\affiliation{Department of Chemistry and Department of Physics \& Astronomy, Rice University, Houston, TX 77005-1892, USA}

\begin{abstract}
Quantum Monte Carlo (QMC) methods have received considerable attention over the last decades due to their great promise for providing a direct solution to the many-body Schrodinger equation in electronic systems. Thanks to their low scaling with number of particles, QMC methods present a compelling competitive alternative for the accurate study of large molecular systems and solid state calculations. In spite of such promise, the method has not permeated the quantum chemistry community broadly, mainly because of the fixed-node error, which can be large and whose control is difficult. In this Perspective, we present a systematic application of large scale multi-determinant expansions in QMC, and report on its impressive performance with first row dimers and the 55 molecules of the G1 test set. We demonstrate the potential of this strategy for systematically reducing the fixed-node error in the wave function and for achieving chemical accuracy in energy predictions. When compared to traditional quantum chemistry methods like MP2, CCSD(T), and various DFT approximations, the QMC results show a marked improvement over all of them. In fact, only the explicitly-correlated CCSD(T) method with a large basis set produces more accurate results. Further developments in trial wave functions and algorithmic improvements appear promising for rendering QMC as the benchmark standard in large electronic systems.

\end{abstract}
\maketitle

\section{Introduction}

Numerical methods have become an invaluable tool in almost all branches of the physical sciences and have been particularly important in the fields of quantum chemistry and solid state physics. For classical systems, direct stochastic simulation techniques such as Monte Carlo and Langevin molecular dynamics have been an essential tool in understanding diverse phenomena ranging from crystalline structures to liquids \cite{FrenkelSmit02,Allen89}.
For quantum systems, the fundamental problem is solving the Schrodinger equation. Unfortunately, exact methods for solving the fermionic Schrodinger equation rely on an explicit construction of the Hilbert space of the problem and thus scale exponentially with system size. Through a combination of faster computers and better algorithms, the prospects for a direct solution to the Schrodinger equation has become realistic for a number of important systems; as of today, however, we must resort to approximate schemes for obtaining results in larger systems of practical interest.  At high temperatures, where there is a continuous crossover between classical and quantum degrees of freedom, Monte Carlo methods in the form of Path Integral Monte Carlo are the dominant approach.  Such methods have been important in the study of hydrogen and other light elements at high pressures and high temperatures \cite{Militzer00}.  It is then surprising that ground state quantum problems of limited size, such as those arising from the electronic degrees of freedom of atoms and small molecules, are not dominated by stochastic approaches.

Traditional quantum chemistry (QC) methods currently offer the best compromise between accuracy and computational cost for molecular ground state calculations. The coupled cluster method with singles, doubles and perturbative triples, CCSD(T), is generally considered the ``gold standard" in the field. It should be pointed out, however, that CCSD(T) is very reliable only for problems dominated by so-called weak correlations that are predominant in the upper rows of the periodic table. In the presence of strong correlations (degeneracy or near degeneracies in the underlying reference determinant), multi-reference CC methods are required \cite{Musial07}, and these are far less developed and limited to relatively small systems. On the other hand, QMC methods, based on a stochastic solution of the Schrodinger equation, offer a promising alternative to traditional QC methods. On the positive side, QMC offers low scaling with particle number yielding efficient solutions for large systems where QC methods are prohibitively expensive. QC methods are generally not size extensive except for CC theory. QMC is not size extensive either but the error can be controlled by working on the Gamma point with periodic boundary conditions (PBC) \footnote{It is also possible to apply Twist Averaged Boundary Conditions \cite{Lin01}, which helps remove the size effects in the energy caused by the quantization of momentum with PBC.} and extrapolating to the thermodynamic limit \cite{Chiesa06,Drummond08}. In addition, QMC has a weak dependence on the basis set, as opposed to QC methods that rely on tailor made basis sets combined with basis set extrapolation techniques. On the negative side, the application of QMC to electronic structure suffers from the well-known fermion sign problem, which results in errors that are difficult to control and predict. In order to produce a practical method, the fixed-node approximation is typically employed, and while the accuracy is only limited by the choice of trial wave function, a simple ansatz often yield results that do not always compare satisfactorily with experiment. A clear approach for improving the accuracy of QMC is simply to improve the quality of the trial wave-function.

Despite their promise, QMC methods have not received a lot of attention in the QC community. Two factors have likely contributed to this state of affairs: the low accuracy of routine QMC calculations when compared to CCSD(T), and the lack of black-box user-friendly computer codes that dominate the QC landscape. In addition, the extension of QMC methods towards the calculation of properties other than ground state energies has been a slow process. Fortunately, this has accelerated over the last years, where applications of QMC methods have appeared in the calculation of excitation energies in large molecules\cite{Send11}, in the structural optimization of molecules including excited states\cite{Barborini12}, and in the calculation of molecular electric properties\cite{Coccia12}, to mention a few.  In this Perspective, we address the first of these two obstacles by using an improved trial wave function.  We have recently developed a scheme for the fast and efficient computation of large multi-determinant expansions in QMC \cite{Clark11}.  This scheme allows us to perform QMC efficiently with thousands of determinants. Being able to perform such large multi-determinant expansions gives us a knob that can be systematically improved as well as an effective parameter that can be extrapolated.  Additionally, the use of an arbitrary bosonic Jastrow factor, implicitly included in QMC through the fixed node approach, makes determinants used in the QMC framework significantly more powerful than the equivalent number of determinants used in a QC method.  We here present benchmark examples where selecting determinants for QMC from a truncation of the second order CI (SOCI) expansion gives results significantly better than SOCI itself, and comparable in quality to CCSD(T)-R12 extrapolated to the complete basis limit. We exemplify the QMC quality improvement originating from this ansatz with calculations on first row molecules.

The organization of this paper is as follows. In section \ref{sec:optm} we describe the form of the wave function used in the calculations, as well as the optimization procedure employed. In section \ref{sec:dimers} we discuss our results on first row atoms and dimers. Finally, section \ref{sec:G1} describes the results on the 55 molecules of the G1 set. This is followed by a discussion of the implications of this work in the QMC community and possible future research.

\section{Trial Wave Function and Optimization Method} 
\label{sec:optm}

There has been significant improvement in QMC methods over the last decade. Some of these improvements have focused on the development and implementation of new wavefunctions including: Pfaffians \cite{Bajdich06},  generalized valence bond (GVB) \cite{Anderson10},  antisymmetrized geminal power (AGP) \cite{Casula04}, backflow transformation \cite{Kwon98}, and multi-determinant expansions \cite{Bajdich10,Seth11,Clark11}. This has been combined with the development of robust and efficient optimization methods \cite{Umrigar07,Toulouse08} and the introduction of novel basis functions (for effective core potentials (ECP) \cite{Petruzielo11} and for all-electron calculations \cite{Esler10}). We recently introduced a computational scheme that allows the quick and efficient evaluation of multi-determinant expansions in QMC calculations \cite{Clark11}, allowing us to routinely use thousands of determinants in calculations with molecules at a small increase in computational cost. In this article, we make use of this method to demonstrate the power of QMC when the trial wave function is systematically expanded in determinants. The wave-function takes the form:
\begin{equation}
\Psi=\exp[-J(R)]\sum_{k}\alpha_{k}\det M_{\uparrow k}\det M_{\downarrow k}
\end{equation}
where $J$ is a Jastrow function which contains one, two, and three body correlation terms, and
 $\alpha_{k}$ is the weight of
the $k'$th determinant configuration. The $(i,j)$ element of matrix $M_{k}$ is
equal to \[
M_{\sigma k}[i,j]=\phi_{j}(r_{i})\]
 where $\phi_{j}$ are 3 dimensional single particle orbitals (s.p.o)
selected from a given orbital set. 

The parameters in the Jastrow and the contraction coefficients $\{\alpha\}$ of the multi-determinant expansion are optimized by the linear optimization method first pioneered by Umrigar and co-workers \cite{Toulouse07}. This method diagonalizes the Hamiltonian in a sub-space formed by taking the derivatives of the trial wave functions with respect to its parameters and linearizing them. In the case of a trial function which is 
linear in the parameterization, such as a multi-determinant expansion with no Jastrow, the resulting sub-space is complete, and the solution to the eigenvalue equations is the lowest energy solution. Rescaling of the lowest energy eigenvector is allowed when non-linear parameters are included, either by a line minimization, controlling the normalization, or any other criterion, to speed the convergence to the minimum. Typically, fewer than ten iterations are required, each using an increasing number of Monte Carlo steps.
    
\section{First Row Dimers}
\label{sec:dimers}

In this section, we use large multi-determinant expansions in all-electron calculations of first row atoms and dimers. There are very accurate total energy calculations on these systems \cite{Chakravorty93, Langhoff91,Linstrom05,Bytautas06,Toulouse08,Bytautas12} and they have traditionally been used to test new wavefunctions in QMC, offering an excellent opportunity to test the power of this methodology. All quantum chemistry calculations presented in this section were performed with the GAMESS code \cite{gamess}, using the Roos augmented triple zeta atomic natural orbital gaussian basis set \cite{Widmark1990,EMSL,Feller96}. In our approach, the ultimate accuracy is currently limited by the choice of molecular orbitals (MO) in the calculations and the selection of determinant configurations included in the expansion. Since we don't currently optimize the single-particle orbitals directly in QMC, we need to start with a reasonable set of MOs to reduce the number of determinants needed to achieve a given accuracy. For the calculations in this section, we use the natural orbitals (NO) of a self-consistent second-order configuration interaction (SOCI) calculation, with an active space including all electrons in 10 orbitals, and up to 40 orbitals in the virtual space. We use configurations state functions (CSF), which are spin and space adapted linear combinations of determinants, and these were selected using a cutoff on the expansion coefficients of the SOCI calculation.   
The Jastrow factor consisted of one, two, and three body terms and all the parameters were optimized simultaneously using the method described in section \ref{sec:optm}.

\begin{table}[th]
\begin{center}
\begin{tabular}{c c c c c c c c}
\hline
\hline
\multicolumn{8}{c}{Atoms} \\
   &    Li ($^2$S) & Be ($^1$S) & B ($^2$P) & C ($^3$P) & N ($^4$S) & O ($^3$P) & F ($^2$P)  \\
\hline
\multicolumn{8}{c}{} \\
$\#$ CSFs & 81 & 160 & 396 & 651 & 755 & 873  & 1051 \\
VMC   &  -7.47766(2) &  -14.66688(4)  &  -24.65248(6)  &  -37.8423(1)  &  -54.5854(2)  &  -75.0620(4)  &  -99.7275(5)  \\
DMC   &  -7.478052(7)  &  -14.66728(2)  &  -24.65359(4)  &   -37.84438(5)  &  -54.58829(7)  &  -75.06591(8)  &  -99.7325(1)  \\
Estm. Exact   & -7.4780603 & -14.66736 & -24.65391 & -37.8450 & -54.5892 & -75.0673 & -99.7339 \\
VMC-corr $\%$   & 99.13(4) & 99.50(4)  & 98.86(5) & 98.25(6) & 97.7(1) & 98.0(2) & 98.0(2) \\
DMC-corr $\%$   &  99.9(2) & 99.92(1) & 99.74(1) & 99.61(1) & 99.52(1) & 99.46(3) & 99.56(3) \\
\ \\
\hline
\multicolumn{8}{c}{Dimers} \\
   &    Li$_2$ ($^1\Sigma_g^{+}$) & Be$_2$ ($^1\Sigma_g^{+}$) & B$_2$ ($^3\Sigma_g^{-}$) & C$_2$ ($^1\Sigma_g^{+}$) & N$_2$ ($^1\Sigma_g^{+}$) & O$_2$ ($^3\Sigma_g^{-}$) & F$_2$ ($^1\Sigma_g^{+}$)  \\
\hline
Bond Length & 5.051 & 4.65 & 3.005 & 2.3481 &  2.075 & 2.283 & 2.668 \\
$\#$ CSFs & 526 & 924 & 2429 & 2937 & 2443 & 3033  & 2537 \\
VMC  & -14.9941(2) &  -29.3363(1)  &  -49.4071(3)  & -75.9108(1)  &  -109.5214(3)  & -150.2991(9)  & -199.498(1)  \\
DMC  & -14.99481(6)  &  -29.33865(6)  &  -49.4131(2)  &  -75.9205(3)  &  -109.5367(3)  &  -150.3194(3)  &  -199.5213(3) \\
VMC-extrap  &  -14.9941(2) &  -29.3370(2)  &  -49.4093(5)  &  -75.9157(3)  & -109.5224(4)  &  -150.305(2)  &  -199.501(1) \\
DMC-extrap  &  -14.99481(6)  &  -29.33872(5)  &  -49.4137(2)  &  -75.9229(6)  &  -109.5372(3)  &  -150.3216(3)  &  -199.5219(3) \\
Estm. exact & -14.995(1) & -29.3380(4) & -49.4141 & -75.9265 & -109.5427 & -150.3274 & -199.5304 \\
VMC-corr $\%$   & 99.3(2)  & 99.5(1) &  98.5(2) &  97.92(6) &  96.30(7)  & 96.6(3)  & 96.1(1)	\\
DMC-corr $\%$   & 99.85(5) &  100.35(2)  & 99.88(6) &   99.3(1)  &  99.00(5)  &  99.12(5) &   98.88(4)  \\
\ \\
\hline
\end{tabular}
\end{center}
\label{Table:atoms_data}
\caption{Summary of QMC results for first row atoms and dimers. Estimates of exact results are taken from \cite{Chakravorty93, Langhoff91,Linstrom05,Bytautas06,Toulouse08,Bytautas12}. }
\end{table}%

Table I shows the calculated VMC and DMC energies for the largest MSD expansions employed in this work, along with the energies extrapolated to zero cutoff and the estimated exact energies. The percentage of correlation energy recovered by the method is also shown, along with the number of CSFs used in the reported calculations. In all cases, the extrapolated DMC energies recover at least $99\%$ of the correlation energy. 

\begin{figure}[ht]
\includegraphics[scale=0.4]{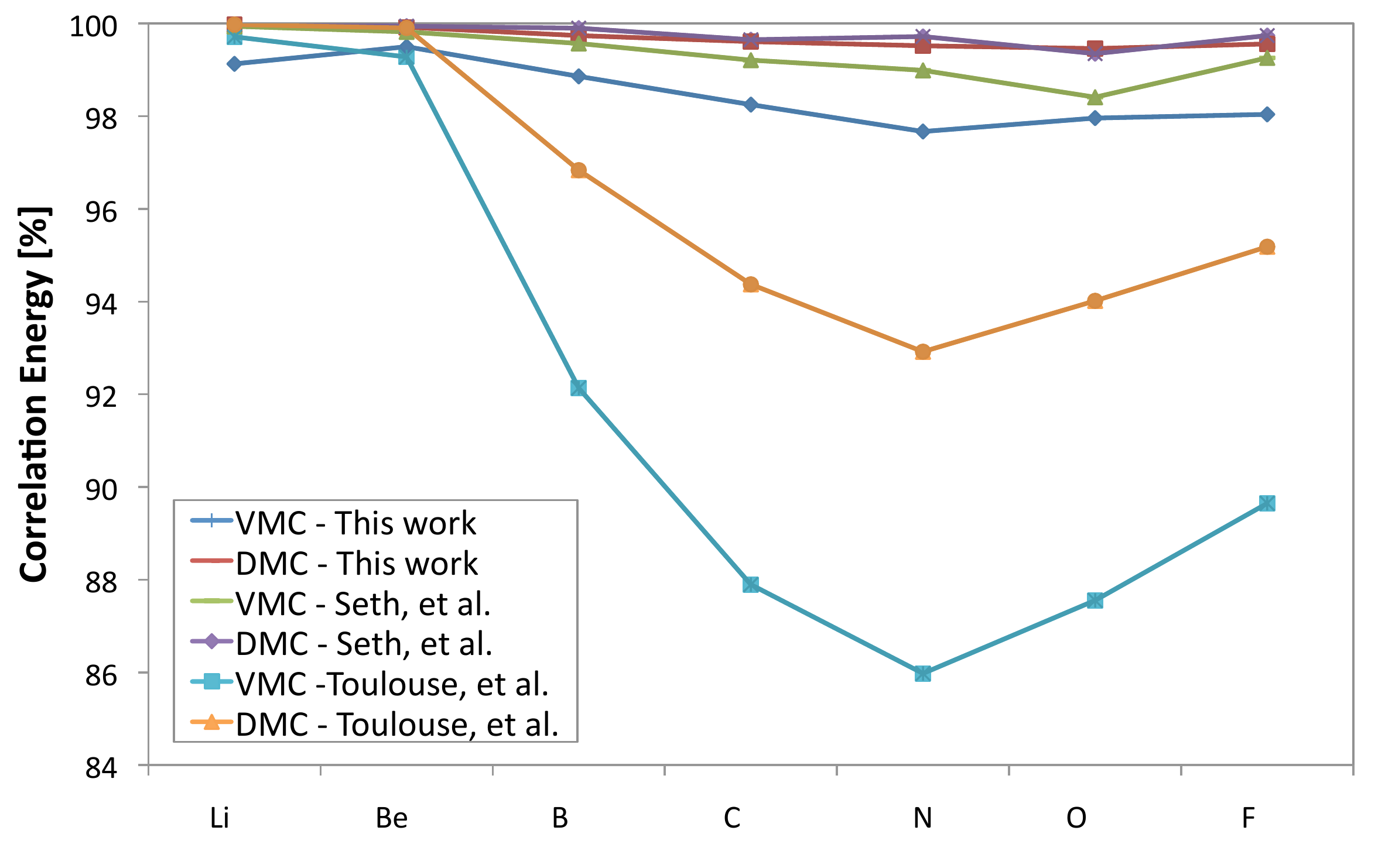}
\label{fig:Atoms_ecor}
\caption{ Percentage of the correlation energy recovered by VMC and DMC methods, for first row atoms. Results reported in this work are compared with recent results from J. Toulouse, {\em et al.} \cite{Toulouse08} and P. Seth, {\em et al.} \cite{Seth11}.}
\end{figure}

Figure \ref{fig:Atoms_ecor} shows a comparison, between our results and several recently published calculations, of the percentage of correlation energy recovered by VMC and DMC for all first row atoms. In the work of Toulouse  {\em et al.} \cite{Toulouse08}, the authors used a MSD expansion similar to the work presented here, but they limited their determinant configurations to those with excitations within the CAS of the valence electrons, or Full Valence CAS (FVCAS). In addition, they perform a full optimization of all the variational parameters in the wavefunction, including molecular orbitals and atomic basis sets (e.g., gaussian exponents). The work of P. Seth, {\em et al.} \cite{Seth11} is also based on small MSD expansions, but with the inclusion of optimized backflow transformations. As can be seen from the comparison, limiting the configurations in the MSD expansion 
to a small FVCAS has a strong effect in the amount of correlation energy recovered by the calculation; excitations to higher virtual states contribute significantly towards the reduction of the remaining fixed-node error. This is even more pronounced when we consider the fact that they do a better optimization of the wavefunction, since they optimize the molecular orbitals at the VMC level whereas we are limited to the orbitals produced by the SOCI method. The inclusion of backflow, on the other hand, produces a large improvement on the results. This is clearly seen by noticing that they used MSD expansions approximately 10-50 times smaller than in this work, yet their VMC energies are typically better than ours, while the DMC energies are similar. 
Although our MSD expansions are at least an order of magnitude larger, 
we anticipate that the computational cost of these different approaches should be competitive as backflow 
transformations are expensive and our MSD expansions are evaluated extremely efficiently. 
Regardless of this, the use of backflow makes a considerable improvement to both energies and variances in QMC, so its use combined with very large MSD expansions should produce very accurate results. This is currently being investigated.

\begin{figure}[ht]
\includegraphics[scale=0.4]{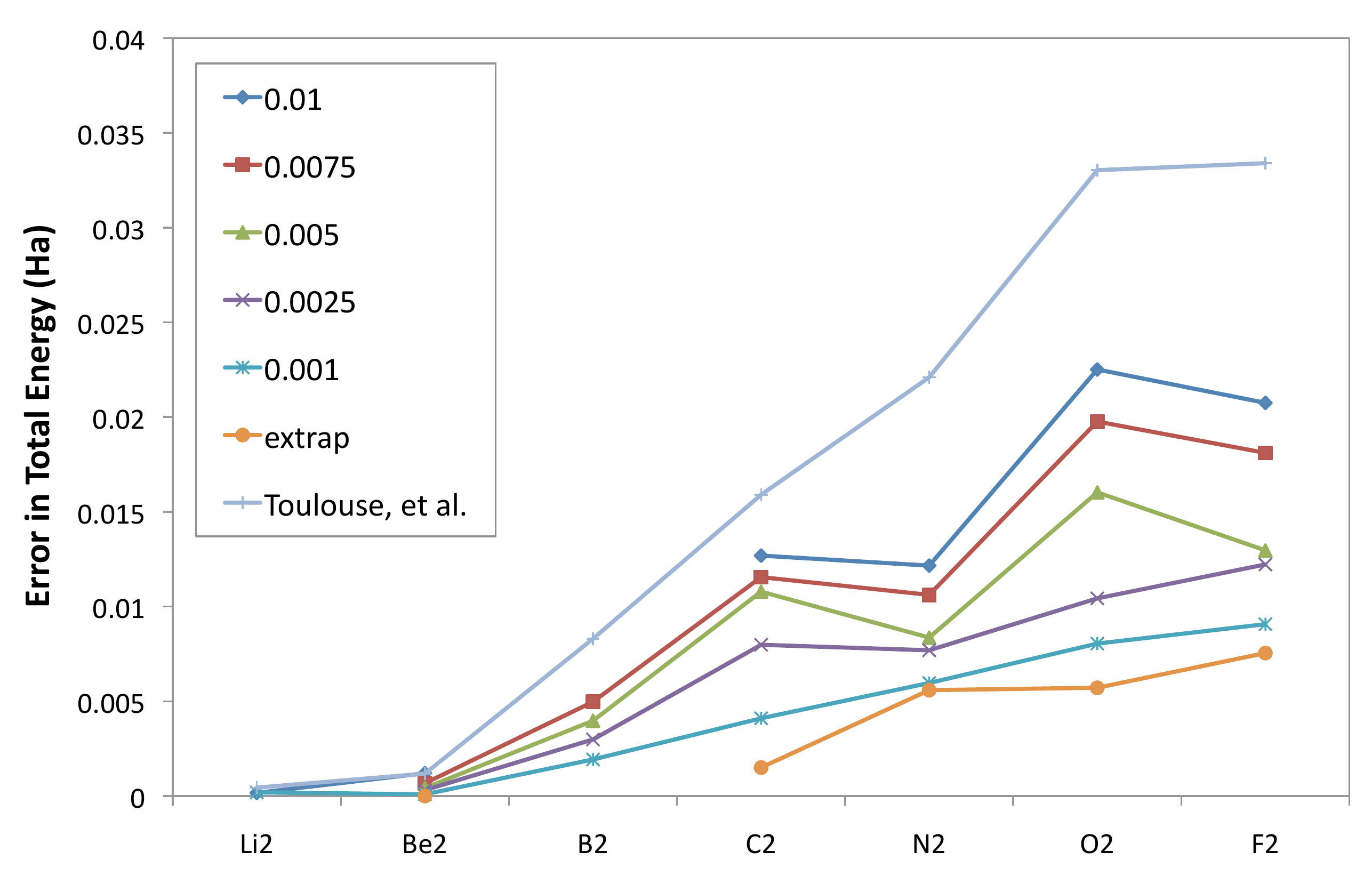}
\label{fig:Atoms_etot}
\caption{ Error in the DMC total energy of first row dimers, as a function of the cutoff in the MSD expansion. Results from J. Toulouse, {\em et al.} \cite{Toulouse08} are also shown for comparison.}
\end{figure}

Figure \ref{fig:Atoms_etot} shows the error in the DMC total energies (relative to the near `exact' results) for all first row dimers, for various cutoffs in the MSD expansion. Results from the FVCAS calculations of Toulouse  {\em et al.} \cite{Toulouse08} are also shown for comparison. Similar to the case of first row atoms, limiting the determinant configurations to valence orbitals only produces total energies with limited accuracy. On the other hand, a systematic expansion into larger MSD sets is clearly seen to converge towards the correct results. The results presented here are the most accurate QMC total energy calculations of first row dimers at equilibrium to date. 

An important point to notice in these results is the increase in the errors with increasing atomic number. There are several reasons for this. On the one hand, the DMC energies have some dependence on the quality of the atomic basis sets in the core region, and this effect gets more pronounced with larger atomic number. This effect can be reduced with better basis sets, tailored for all-electron QMC calculations, but we have not pursued this in this work. On the other hand, in this section we limit the determinant configurations to single and double excitations outside the employed CAS (SOCI); this is not size extensive and has a stronger effect in systems with more electrons. Nonetheless, we clearly show the capacity of large multi-determinant expansions to capture a large fraction of the correlation energy missed at the single determinant level, reaching high levels of accuracy and systematically reducing the fixed-node error.    

\section{G1 set}
\label{sec:G1}
In this section, we apply the MSD wave-function to the 55 molecules in the G1 set \cite{Curtiss90,Grossman02}. We have used the same geometries as in previous QMC studies \cite{Nemec10}, in order to have a clear comparison with previous results. With the exception of hydrogen atoms, we used the Burkatzi-Filippi-Dolg (BFD) set of pseudopotentials and corresponding optimized basis sets \cite{Burkatzki07}. 
The Coulomb potential was used for hydrogen with the Roos augmented triple zeta atomic natural orbital gaussian basis set \cite{Widmark1990,EMSL,Feller96}. We have used the MP2 natural orbitals (NO) in the multi-determinant expansion and the determinant configurations were chosen from CISDTQ calculations with those orbitals. Similar to the previous section, we use configuration state functions to eliminate redundant
variational parameters from the wave function and facilitate the optimization process; the number of configurations is controlled by applying a cutoff to the resulting CISDTQ wave-function based on the magnitude of the expansion coefficient. To this multi-determinant expansion we add a Jastrow factor that contains 1,2, and 3 body terms and optimize all the variational parameters simultaneously (including the nonlinear Jastrow coefficients and the linear CSF coefficients) with the method described in section \ref{sec:optm}. DMC calculations were performed with a timestep of 0.0025 $Ha^{-1}$ using a target population of $\approx 2500$ walkers. In addition, we used T-moves \cite{Casula06} to obtain a rigorous upper bound to the ground state energies.

Throughout this section, we will use results from explicitly-correlated CCSD(T) calculations extrapolated to the complete basis set limit (CBS-CCSD(T)-F12) as a reference. These calculations were performed with the MOLPRO software package \cite{molpro_long,Hampel92,Knowles93,Deegan94}. The F12 calculations \cite{Alder07,Knizia08, Knizia09} were performed with density fitting \cite{Manby03} and resolution of the identity with optimized basis sets \cite{Weigend08,Peterson08,Yousaf09_1,Yousaf09_2}. This represents the gold standard in quantum chemistry for the types of molecules here analyzed and has been shown to produce very accurate results in the study of sets of molecules similar to the one we study here \cite{Feller08}. We performed the basis set extrapolation based on the VTZ, VQZ and V5Z basis sets; the errors in the extrapolation procedure are expected to be on the order of 0.1 mHa \cite{Feller11}. As shown by Feller, {\emph et al.} \cite{Feller08}, higher order corrections to the FC-correlation energy are typically on the order of 0.2-0.5 mHa, although there are a few notable cases in the molecules studied in this work like CN with a correction of $\sim$1.57 mHa, and molecules like $ClO$, $HNO$, $Si_2$, $O_2$, $F_2$, and $P_2$ with corrections around $\sim$1 mHa. These corrections are likely to be smaller in our case, since we use pseudopotentials.

\begin{figure}[t]
\includegraphics[scale=0.37]{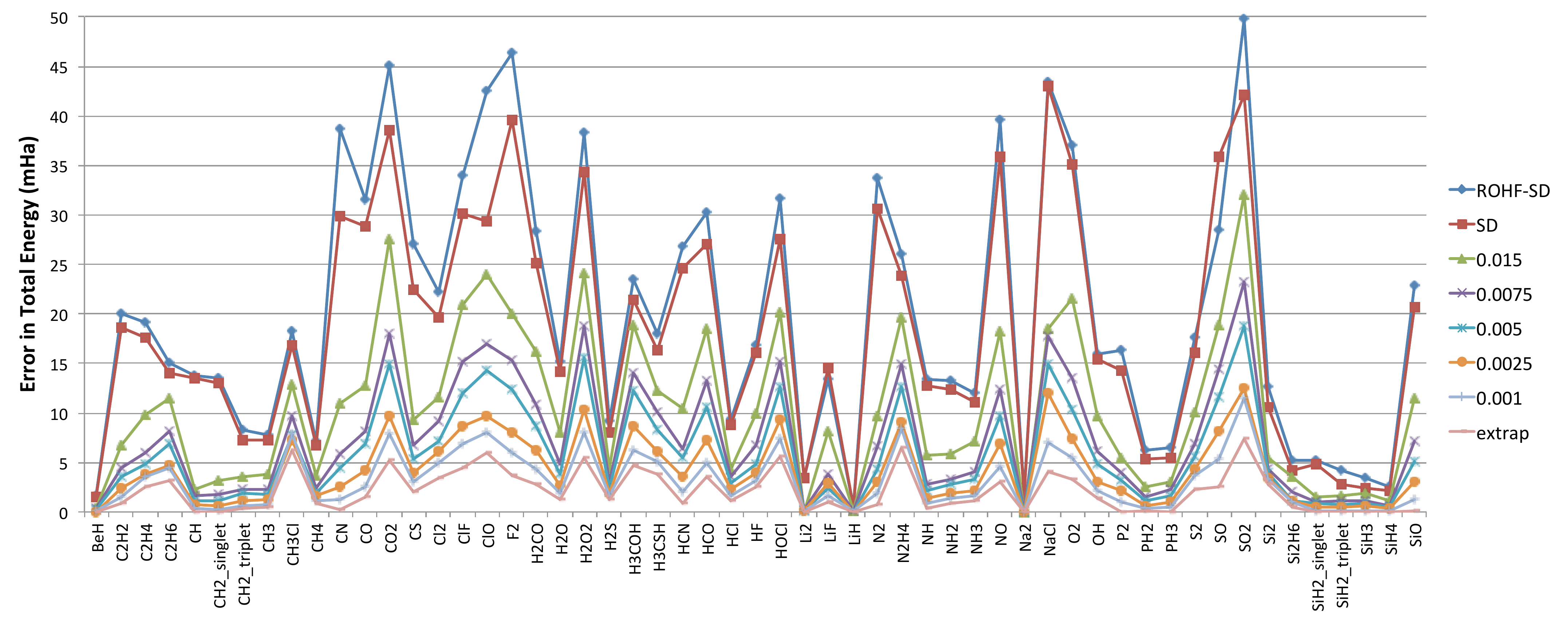}
\label{fig:Error_ET}
\caption{DMC total energy of all 55 molecules of the G1 set considered in this work, for various cutoffs in the MSD expansion. The figure also includes the energy obtained by extrapolating the results to zero cutoff. }
\end{figure}

Figure \ref{fig:Error_ET} shows a summary of the difference in total energies between DMC and CBS-CCSD(T)-F12 for all 55 molecules in the G1 set. Various cutoffs in the MSD expansion are shown, as well as single determinant cases both with ROHF orbitals and MP2-NO. The figure also includes the energy obtained by extrapolating the results to zero cutoff \cite{Bajdich10,Clark11}. We see a dramatic improvement in the DMC energies with decreasing cutoff; in fact the mean absolute error (MAE) decreases from $\sim18$ mHa in the single determinant case, to $\sim3$ mHa with a cutoff of 0.001, and to $\sim2$ mHa for the extrapolated values. This is more pronounced in the case of difficult molecules like SO$_2$ where the MSD expansion recovers $\sim35$ mHa with respect to the single determinant case; this represents $\sim88\%$ of the fixed-node error. The dependence of the energy with the number of configurations in the expansion varies across the set. This is not unexpected since it will in general depend on the multi-configurational character of the molecule and on the ability of MP2 to produce an accurate orbital set. This is clearly shown when you compare molecules like $CH_{n}$ with molecules like $SO_2$ and $H_{2}O_{2}$. In the former, the inclusion of a few configurations is enough to recover most of the fixed-node error while in the latter the improvement is slow and more systematic.   

\begin{figure}[t]
\includegraphics[scale=0.55]{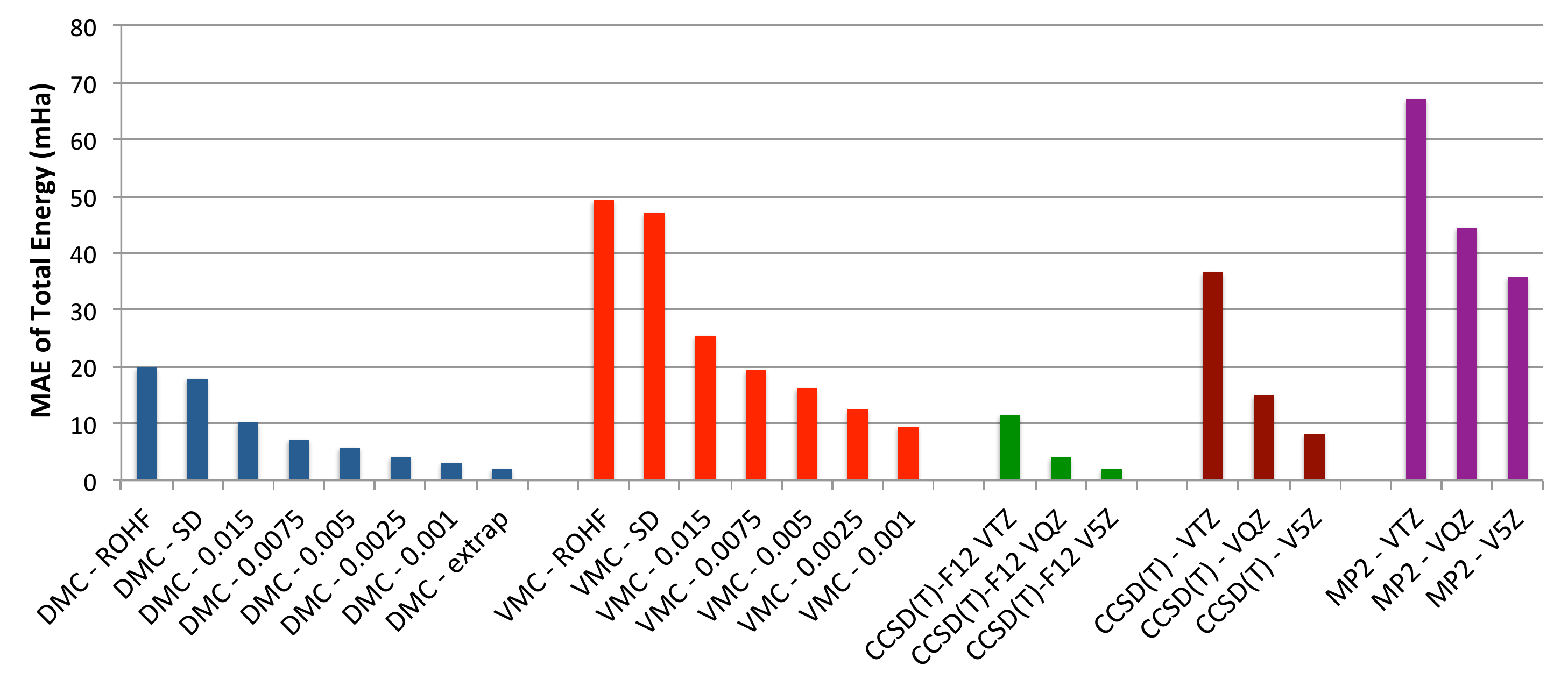}
\caption{Comparison of the mean absolute error (MAE) of the total energy between QMC and several traditional quantum chemistry methods.   }
\label{fig:MAE_ET}
\end{figure}

\begin{figure}[th]
\includegraphics[scale=0.5]{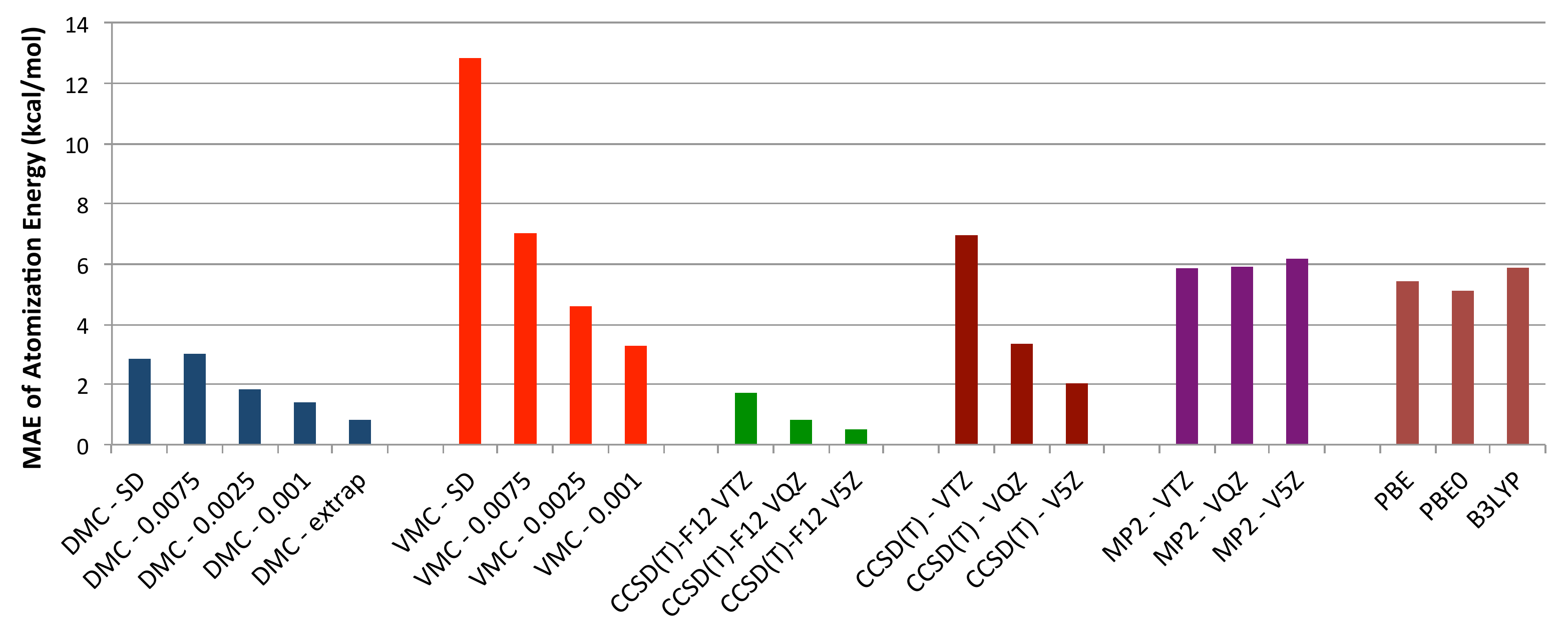}
\caption{Comparison of the mean absolute error (MAE) of the atomization energies between QMC and several traditional quantum chemistry methods.  }
\label{fig:MAE_AE}
\end{figure}

Figures \ref{fig:MAE_ET}  and \ref{fig:MAE_AE} show the MAE in the total energy and the atomization energy respectively, always taking CBS-CCSD(T)-F12 as reference. In both cases we show the dependence with cutoff in the MSD expansion, as well as results from various standard quantum chemistry methods, in particular MP2, CCSD(T), CCSD(T)-F12 and DFT with various functionals.  Only the CCSD(T)-F12 method 
is able to produce results that are better than DMC. Traditional quantum chemistry methods must be combined with large tailor-made atomic basis sets in order to reach chemically accurate results; DMC results, on the other hand, have a much smaller dependence on basis sets.   

These results are very encouraging for several reasons. The computational cost of traditional quantum chemistry methods has a very steep scaling with system size, e.g. MP2 scales as $N^5$, CCSD as $N^6$ , and CCSD(T) as $N^7$, where N is representative of the size of the system. DMC, on the other hand, scales as $N^3$ for the energy per electron in its basic implementation which can be improved to $N^2$ with efficient wave function evaluation techniques and possibly to linear scaling [3,4]. While it is unlikely that QMC replaces CCSD(T) as the standard method in the study of small molecular systems (up to 5-10 atoms), these results show that it has the potential of becoming the community standard in large molecular systems and periodic calculations. In addition, the results on the G1 set show the great promise of new wave functions in QMC. While large multi-determinant expansions clearly show remarkable potential, it is possible to obtain even better results with the inclusion of more elaborate choices like a systematic expansion in geminals [5], and multi-determinants combined with backflow [6], to name a few. Finally, the use of large determinant expansions has great potential in the study of low lying excited states, offering a natural basis set in the Correlated Function Monte Carlo method [7]. We expect a dramatic improvement in the study of excitation energies of molecular systems with this method, with a possible extension to periodic systems and solids with the use of localized molecular orbitals.

\begin{figure}[t]
\includegraphics[scale=0.5]{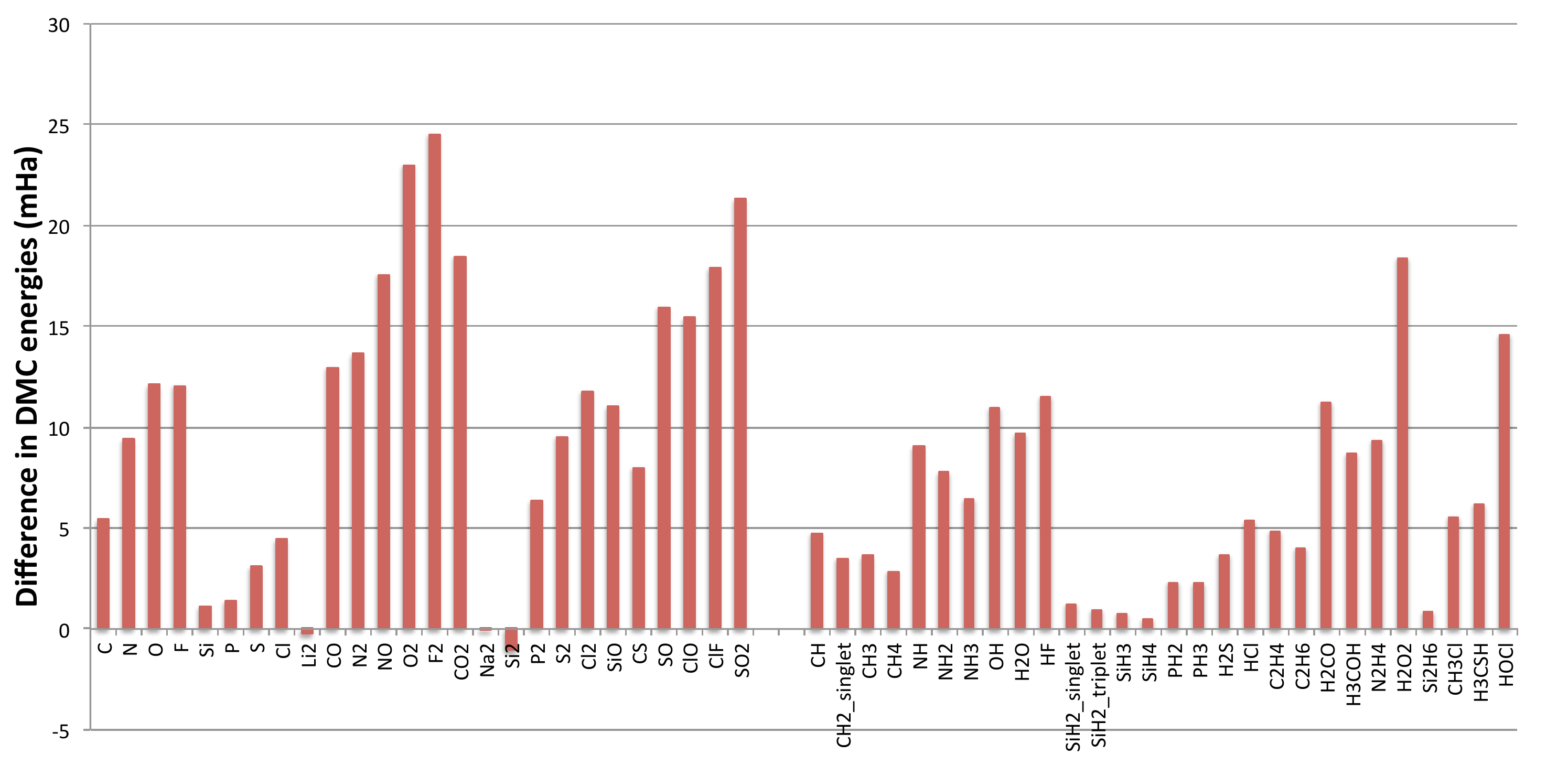}
\caption{Comparison of the fixed-node DMC energies between this work and the recent calculations of F. R. Petruzielo, {\emph et al.} \cite{Petruzielo12}. Their calculations were performed with an ECP for hydrogen, as opposed to our calculations where the Coulomb potential was used. Although the effect should be small, the comparison in the case of molecules containing hydrogen (moved to the right end of the figure) will be affected by this and should be taken with care.   }
\label{fig:comparison_Cyrus}
\end{figure}

Recently, calculations of the atomization energies on the G1 set with QMC methods have been reported by F. R. Petruzielo, {\emph et al.} \cite{Petruzielo12}. In their work, they also used the BFD set of pseudopotentials and optimized basis sets, with the exception of hydrogen where they used an optimized pseudopotential created for their work. While they also used multi-determinant expansions in their work, they limited the excitations in the determinant to the FVCAS of the molecules. Figure \ref{fig:comparison_Cyrus} shows the difference between the best DMC energies reported by F. R. Petruzielo, {\emph et al.} and the DMC energies with the largest multi-determinant expansion used in this work. Notice that the hydrogen atoms are described in different ways in both sets of calculations, so comparisons should be made with care. They report a MAE in the atomization energies (comparing to experimental results) of 1.2 kcal/mol, in good agreement with our best results. Notice that this is mainly due to error cancelation between the energies of the molecules and the corresponding atoms. This suggests that, while large multi-determinant expansions are necessary in order to systematically reduced the fixed-node error in the DMC energies, a scheme to introduce size-consistency is also required to produce accurate atomization energies. This will be particularly important in the study of larger molecular systems with this approach.

\section{Conclusions}

In this Perspective, we have shown that QMC with our improved multi-Slater-Jastrow wave function gives results on atoms and small molecules that are competitive with the best QC predictions.  In particular, for first row dimers and the molecules in the G1 set, we find that QMC is strictly more accurate than MP2, CCSD(T), various DFT approximations, and even competitive with CCSD(T)-R12.   For first row atoms and dimers, QMC recovers more than 99\% of the correlation energy in all cases.  For the G1 set, the MAE in total energy is approximately 2 mHa and the MAE in atomization energies is approximately 0.8 kcal/mol. These results demonstrate that QMC is a promising alternative to traditional QC methods for a wide variety of systems.

The particular ansatz that we use in this paper has the added advantage of being systematically improvable.  Increasing the number of determinants yields improved results. This means (1) that one can extrapolate based on this parameter, (2) that there is a systematic way to improve accuracy if a better result is desired, and (3) that one can estimate whether the answer has converged. These points, combined with the variational upper bound that comes from QMC, gives significantly more control over the fixed node error than has been previously feasible from single shot QMC calculations.  In fact, we believe that the results presented in this paper are already a good database for understanding the fixed node error on the single determinant Slater-Jastrow wave function as it relates to different molecular and atomic systems.

Finally, although we find the current QMC status already impressive, we believe that this may be the first step in a longer quest for approaching chemical accuracy on larger molecules and condensed systems.  Beyond the results shown in this article, our optimism arises from two facts.  First, QMC is able to effectively leverage increases in computational power as a result of its natural parallelization and low scaling with particle number.  Second, research into new variational ansatz is still in its infancy and we speculate that further improvements as equally impressive as those reported here may be achievable. Our optimism suggests that QMC is a fertile research field worthy of receiving more attention.

\begin{acknowledgments}
This work was performed in part under the auspices of the US Department of Energy (DOE) by LLNL
under Contract DE-AC52-07NA27344. 
The work at Rice University was supported by DOE DE-FG02-04ER15523 and the Welch Foundation (C-0036).
MAM and GES also acknowledge support from CMCSN award DE-FG02-11ER16257.
The work at UI was supported by the National Science Foundation under No.
0904572 and EFRC - Center for Defect Physics sponsored by the US DOE, Office of Basic Energy Sciences (BES)
and JK by DOE-BES Materials Sciences and Engineering Division.

\end{acknowledgments}

\end{document}